\documentclass{article}
\usepackage{amssymb}
\usepackage[nobysame]{amsrefs}

\usepackage{graphicx}
\usepackage{amsmath}
\usepackage{epsfig}
\usepackage{color}

\newtheorem{theorem}{Theorem}

\newtheorem{definition}[theorem]{Definition}

\begin{document}
 \title{Laplace's law for  sharp and diffuse interfaces}

	\author{ Sergey  Gavrilyuk\,\, and \, Henri Gouin \thanks{E-mails:\newline
		 	sergey.gavrilyuk@univ-amu.fr; \,\  \,
				henri.gouin@univ-amu.fr; henri.gouin@ens-lyon.org} }
	\date{\footnotesize Aix--Marseille University,
		IUSTI,  CNRS UMR 7343, Marseille, France.}
	\maketitle

\begin{abstract}
 We study both  diffuse and  sharp liquid--vapor  interfaces.  The equilibrium equation of  fluids is derived by using the principle of virtual work in a  domain including the  interfaces. 
 For diffuse interfaces, the  surface tension coefficient depends on   the  density profile across the interface. 
 For sharp interfaces, the liquid--vapor  layer is mathematically represented by a geometric surface and   its specific energy   is a Dirac delta function at the surface. 
 
 We compare the both approaches and find relations between the surface tension coefficient  and parameters of the models. 
\begin{center}
{\bf Résumé}
 \end{center}
 
Nous étudions à la fois les interfaces  diffuses et celles sans épaisseur dans un domaine liquide--vapeur. L'équation d'équilibre des fluides est obtenue à l'aide du principe du travail virtuel dans un domaine incluant les interfaces. 
 Pour les interfaces diffuses, le coefficient de tension superficielle dépend du profil de densité à travers l'interface. 
Pour les interfaces sans épaisseur, la couche liquide--vapeur est représentée mathématiquement par une surface géométrique et son énergie spécifique  est une fonction delta de Dirac à la surface. 
 
 Nous comparons les deux approches et établissons des relations entre le coefficient de tension superficielle et les paramètres des modèles. 
 
\end{abstract}
 
 \section{Introduction}
 Phase separation between liquid and vapor can  mathematically be explained by the fact that
 the  internal energy per unit volume  of
 homogeneous fluids is a non-convex function of   density   and  entropy. For example at a given temperature, this
 non--convexity property is related with  non--monotonicity of
 the thermodynamic  pressure.

 In continuum  mechanics the simplest model for describing
 inhomogeneous fluids inside interfacial layers considers the 
energy density  as the sum of two terms: the first
 one  
 corresponds to a homogeneous fluid defined  by a  non-convex equation of state,  and the second one  is associated with the non--uniformity
 of the fluid. This non--uniformity  is approximated by a gradient density expansion at the second order taking account of the square of the density gradient  \cite{Rowlinson_Widom_2013,Vanderwaals_1895}. This form of internal energy 
 can be deduced from molecular mean-field theories where the
 molecules are modeled as hard spheres
 submitted to Lennard--Jones potentials \cite{Domb_1997,Evans_1979}.
 This expression of energy has been introduced by  van der Waals   and
 is now widely used in the literature as \textit{energy of diffuse interface} \cite{Korteweg_1901,Casal_1963,Truskinovskii_1983,Casal_Gouin_1985}.  The idea of van der Waals was further implemented  by Cahn and Hilliard  who proposed a nonlinear   equation of diffusion  for an order  parameter  describing  interfaces
 as \textit{diffuse layers} \cite{Cahn_Hilliard_1958}.
 \\
 It has been extended to more complex situations e.g. for modelling 
 fluid mixtures, porous materials and other  strongly inhomogeneous media as in  \cite{Gouin_Ruggeri_2005,Forest_2009}.
 In the description of  fluid behavior near the thermodynamic critical point,  Rowlinson and Widom 
 said that
 near the critical point, a gradient expansion 
 	truncated in second order in gradient of density, is most likely to be successful and
 	perhaps even quantitatively accurate ({\cite{Rowlinson_Widom_2013}, Chapter 9}).

 This is not the case for interfaces when the fluid is well below its thermodynamic critical point. Experimentally, the interface appears as a region with a thickness on the order of a manometer and cannot be regarded as a diffuse layer of density. We conceive that this transition of liquid to vapor constitutes \textit{a sharp interface}, which  at the very thin  scale, is represented by a density jump: the density behaves like a  Heaviside function, and it is well known that classical methods of diffuse interfaces  fail to describe this type of phenomenon \cite{Lowengrub_1998,Brackbill_1992}. In the limit of   well--defined interface layers,
  the system of equations converges
 		to a   \textit{sharp interface model}. The classical fluid equations are then set on either side of the interface, and jump conditions are imposed at a discontinuity geometric surface \cite{Hou_1994}. Surface tension manifests as a localized \textit{surface force} acting on fluid elements at the  interface \cite{Israelachvili_2011}. The sharp interface approach was also successfully used in multi-phase modeling \cite{Gavrilyuk_2017}.  Because the  density gradient behaves as  \textit{a Dirac delta-function} and  yields an infinite surface energy, the description of the internal energy cannot be described by adding  quadratic  terms in density gradient. \\
The internal energy—whilst it can still be modeled as the sum of two terms—is such that the first term always
corresponds to the fluid having a uniform density equal to its
local composition, but  to avoid the singularity,  the second  term must be limited   to the first order in density gradient. \\

The aim of this note is to understand   consequences of the two above mentioned  models   to  represent Laplace's law of surface tension. 
 In the two cases, it is possible to express  the surface tension  coefficient  involved in  Laplace's law in terms of parameters of the model.  For diffuse interfaces the surface tension  coefficient appears as an integral along the line that is normal to    \textit{thick  surface}   in opposition to  sharp interfaces where the surface tension coefficient is  associated with the difference of the squares of densities between liquid and vapor bulks. \\

 	\section{Continuum in equilibrium}
 	Geometrically, the position of a fluid in equilibrium is  governed by an application
 	$\boldsymbol\varphi$ from $\mathcal{D}_0$ into $\mathcal{D}$, where $\mathcal{D}_0$  and is    $\mathcal{D}$ are   open domains of the three--dimensional  space occupied by the medium. Subscript $^T$ denotes the transposition,  $\boldsymbol X =(X^1, X^2, X^3)^T$ is the reference position in Lagrange variables, and  $\boldsymbol x =(x^1, x^2, x^3)^T$ is the particle position in Euler variables. Usually,  we parameterize  the actual position  of the medium as
\begin{equation*}
	\boldsymbol x=\boldsymbol\varphi(\boldsymbol X),
	\label{flow}
\end{equation*}
where $\boldsymbol X\in\mathcal{D}_0\longrightarrow \boldsymbol x=\varphi(\boldsymbol X)\in\mathcal{D}$ is a differentiable  mapping \cite{Serrin_1959,Gavrilyuk_Gouin_1999}.
\begin{definition} 
	Consider a differentiable application ${\boldsymbol\Phi}$ such that
	\begin{equation*}
		(\boldsymbol X, \varepsilon)\in \mathcal D_0 \times {\mathcal O} \longrightarrow\boldsymbol  x=\boldsymbol\Phi(\boldsymbol X, \varepsilon)\in \mathcal D\quad {\rm with}\quad \boldsymbol\Phi(\boldsymbol X, 0) = \boldsymbol\varphi(\boldsymbol X) \label{virtual motion family}
	\end{equation*}
	where ${\mathcal O} $ is an open interval   of $\mathbb R$ containing zero. We call $\boldsymbol\Phi$ a one-parameter family of \textit{virtual positions},  the real position $x = \boldsymbol\varphi(\boldsymbol X)  
	$ is obtained when $\varepsilon =0$.\\
	The associated \textit{virtual displacement} $\delta\boldsymbol{x}(\boldsymbol X)$ is defined as 
	\begin{equation}
		\delta\boldsymbol{x}(\boldsymbol X)=\frac{\partial \boldsymbol\Phi (\boldsymbol X, 0)}{\partial \varepsilon}.\label{zeta1}
	\end{equation}
\end{definition}
We associate the field of tangent vectors to
$\mathcal{D}$ as   
\begin{equation}
	{\boldsymbol{x}}\in \mathcal{D}\;\  \longrightarrow\;\ {\boldsymbol{\xi}}({\boldsymbol x}) =\delta\boldsymbol{x} \left(\boldsymbol{\boldsymbol\varphi}^{-1} (\boldsymbol x)\right)\in T_{\boldsymbol{x}}(\mathcal{D}),
	\label{zeta2}
\end{equation}
where $T_{\boldsymbol{x}}(\mathcal{D})$ is the tangent space  to $\mathcal{D}$ at
$\boldsymbol{x}$.   
We note that   $	\delta\boldsymbol{x}(\boldsymbol X)$ is a presentation of virtual displacements in Lagrangian coordinates by \eqref{zeta1}, while   $ {\boldsymbol \xi(\boldsymbol x)}$ is the corresponding representation on in Eulerian coordinates by
means of \eqref{zeta2}. Let us note that the analog of virtual displacements in the theory of distributions is  the space of test functions with compact support \cite{Schwartz_1966}. \\
Any physical quantity $f$ (density, velocity, pressure, ...)  can be considered   either in the Lagrangian 
or in Eulerian coordinates 
\begin{equation*}
	\boldsymbol{X}\longrightarrow f(\boldsymbol{X})\qquad
	{\rm or }\qquad
	\boldsymbol{x}  \longrightarrow f(\boldsymbol{x}).
\end{equation*}
For the sake of simplicity, we abuse the notation by using the same letter $f$ both for  Lagrangian and Eulerian representations.  The first  one is obtained from the second   one by replacing $\boldsymbol{x}$ with ${\boldsymbol\varphi}(\boldsymbol{X})$.
\\
Below, symbols
$(q_1, q_2, q_3)$ denote the orthogonal curvilinear coordinates used to locate points of $\mathcal D$ and  $\boldsymbol e_1, \boldsymbol e_2, \boldsymbol e_3$ represent the unit vectors of the coordinate system.  The elementary displacement of a point $\boldsymbol x$ is 
\begin{equation*}
	d\boldsymbol x =\sum_{i=1}^{3}\frac{\partial \boldsymbol x}{\partial q_i} dq_i= \sum_{i=1}^{3} h_i {\boldsymbol e}_i\, dq_i, 
\end{equation*}
where $h_1, h_2, h_3$ are Lam\'{e} coefficients that are continuously derivatiable  functions of  $q_1, q_2, q_3$.
 
 \section{Surface tension of diffuse curved interfaces in equilibrium}

We consider interfaces between two phases of a fluid (as an interface between liquid and vapor) \cite{Casal_Gouin_1985,Rowlinson_Widom_2013}.
Surfaces of equal density materializing the interface are presented as  parallel surfaces in the orthogonal curvilinear coordinates when   fluid density $\rho$ is a function of $q_3$ only: $\rho=\rho(q_3)$ \cite{Rocard_1967}.   Index $3$ refers to the  normal direction to the surfaces of equal density in the direction of increasing density.  Then, the normal unit vector in the direction of $q_3$  is ${\boldsymbol e}_3$. We also denote ${\boldsymbol e}_3={\boldsymbol n}$. We suppose that  the unit normal vector field $\boldsymbol{n}$ is locally extended in the vicinity of $S$. In particular, it implies the expression of the sum of  principal curvatures of parallel surfaces 
\begin{equation*}
	H_s=-{\rm div}\, {\boldsymbol n}.\label{Curvature} 
\end{equation*}
 Neglecting external forces (as gravity forces), we recall that the equation  of equilibrium  are classically issued from the minimization of the potential energy of all the fluid
 \begin{equation*}
 \mathcal E=	\int_{\mathcal D}W (\rho,\nabla\rho ) \,dv=\int_{\mathcal D}\rho\,\alpha (\rho,\nabla\rho ) \,dv
 \end{equation*}
 where $\rho$ is the fluid density and $dv$ the volume element of $\mathcal D$, and in the simplest case of diffuse interfaces the specific (per unit mass) energy is  \cite{Rocard_1967,Casal_1963} 
 \begin{equation}
 \alpha(\rho, \nabla\rho )=\alpha^h (\rho)+\frac{\lambda}{2\rho}\,\vert\nabla\rho\vert^2,  \label{energydiffuse}
 \end{equation}
where $\lambda$ is a positive coefficient depending of the density.
We omit   the dependence of the internal energy $\alpha^h$ on the entropy (or temperature).  In the case of constant temperature,  the internal energy  should be  replaced by the free energy.
Equation of the minimum of energy $\mathcal E$ is 
 \begin{equation}
 	\nabla\left(\frac{\delta W}{\delta \rho}\right)=0, \label{chpot}
 \end{equation}
and classically given in the literature \cite{Casal_Gouin_1985,Rocard_1967,Dunn_Serrin_1985,Casal_1963,Vanderwaals_1895}.\\
Term $\displaystyle \frac{\delta W}{\delta \rho}$ denotes the variational derivative of $W$ and represents the \textit{extended chemical potential for capillary fluids} \cite{Gouin_1987}. Relation expresses \eqref{chpot} that the  extended chemical potential is constant in all the fluid. From equation of state of the thermodynamic pressure  $\displaystyle\mathcal{P}=\rho^2\frac{\partial \alpha^h (\rho)}{\partial \rho}$, we obtain the equilibrium condition in the form   
 \begin{equation}
 	\frac{1}{\rho }	\nabla{\mathcal{P}}-\nabla\left(\lambda\, \Delta\rho\right)=0.
 	\label{equilibrium_equation_diffuse_interface}
 \end{equation}
 Relation \eqref{equilibrium_equation_diffuse_interface} should be satisfied inside  the  domain between the bulks.  \\
From $\displaystyle	\nabla\rho =\frac{1}{h_3}\,\frac{\partial\rho}{\partial q_3}\,\boldsymbol e_3$, the equation in the normal direction is    
 \begin{equation*}
 	\frac{1}{h_3}\,\frac{\partial {\it\mathcal P}}{\partial q_3} =\rho\,	\frac{1}{h_3}\frac{\partial(\lambda\, \Delta \rho)}{\partial q_3}, 
 	\end{equation*}
 	or
 	 \begin{equation}
 	\frac{\partial {\it\mathcal P}}{\partial q_3} =\rho\,\frac{\partial (\lambda\,\Delta \rho)}{\partial q_3}. \label{normal_component}
 \end{equation} 
 \begin{equation*}
{\rm From}\quad \Delta\rho= {\rm div(\nabla\rho) }={\rm div}\left(\frac{1}{h_3}\,\frac{\partial\rho}{\partial  q_3}\,\boldsymbol e_3\right) = \left({\rm div}\, {\boldsymbol e}_3\right)\,\frac{1}{h_3}\,\frac{\partial\rho}{\partial q_3}+\nabla\left(\frac{1}{h_3}\,\frac{\partial\rho}{\partial q_3}\right)^T {\boldsymbol e}_3, 
 \end{equation*}
 we obtain the expression of the Laplacian operator \cite{Germaina_1973}
 \begin{equation*}
 	\Delta\rho= -H_s\,\frac{1}{h_3}\,\frac{\partial\rho}{\partial q_3}+\frac{1}{h_3}\,\frac{\partial}{\partial q_3}\left(\frac{1}{h_3}\,\frac{\partial\rho}{\partial q_3}\right).
 \end{equation*}
We consider the case where $\lambda$ is assumed constant.
 Integrating  equation \eqref{normal_component} with respect to $q_3$, we obtain
 \begin{equation*}
 	{\it\mathcal P}-{\it\mathcal P_v}=\Big[ \lambda\, \rho\, \Delta\rho\Big]_{q_3^v}^{q_3^i}-   \int_{q_3^v}^{q_3^i} \lambda\,\Delta \rho\, \frac{\partial   \rho}{\partial q_3}\, dq_3 
 \end{equation*}
 \begin{equation*}
 	=\Big[ \lambda\, \rho\, \Delta\rho\Big]_{q_3^v}^{q_3^i}- \lambda \int_{q_3^v}^{q_3^i} \left( \frac{1}{h_3}\left(-H_s\,\frac{\partial\rho}{\partial x_3}+\frac{\partial}{\partial q_3}\left(\frac{1}{h_3}\frac{\partial\rho}{\partial q_3}\right)\right)\right)\, \frac{\partial   \rho}{\partial q_3}\, dq_3,
 \end{equation*}
   $q_3^v$ is a position in  a  vapor bulk,  and $q_3^i$ a position  inside the interface.
 Then,
 \begin{equation*}
 	{\it\mathcal P}-{\it\mathcal P_v}= \lambda\Big[\rho\, \Delta\rho-\frac{1}{2}\left(\nabla\, \rho\right)^2\Big]_{q_3^v}^{q_3^i} + \lambda\int_{q_3^v}^{q_3^i}H_s\left(\nabla\, \rho\right)^2  \, h_3\, dq_3,\,\ {\rm where}\,\ {\nabla}\, \rho=\frac{1}{h_3}\,\frac{\partial\rho}{\partial q_3}\,\boldsymbol e_3.
 \end{equation*}
 In the  bulks  all derivatives of the density vanish. Hence, one gets   
 \begin{equation*}
 	{\it\mathcal P_\ell}-{\it\mathcal P_v}=  \lambda\int_{x_3^v}^{x_3^{\ell}}H_s\left(\nabla\, \rho\right)^2  \,  dx_3, 
 \end{equation*}
 where  $dx_3 = h_3\, dq_3$,  and indexes  $\ell$ and $v$ indicate the positions inside liquid and vapor bulks. 
 If the mean curvature $ H_s $ varies slowly and can be considered as a constant through the interface, one gets the classical Laplace law
 \begin{equation}
  \mathcal{P}_\ell-\mathcal{P}_v= \sigma \, H_s,
   \label{Laplace_law_1}
   	\end{equation}
   	 where the surface tension coefficient $\sigma$ is 
 \begin{equation}
 	\sigma =  \lambda\int_{x_3^v}^{x_3^{\ell}}\left(\nabla\, \rho\right)^2  \,  dx_3. 
 	\label{definition1_sigma}
 \end{equation}
The hypothesis $H_s\approx const$ is valid because the capillary layer thickness is measured in $Angstr\ddot{o}ms$, and the radii of curvature of interfaces has non-molecular length \cite{Degennes_1985}.
 Then, from the surface tension value, relation \eqref{definition1_sigma} 	can be considered as the definition  of parameter  $ \lambda$.

   \section{Surface tension of a sharp curved interface}
   The capillary specific  energy is now expressed as\;\ $\mu\, \vert \nabla \rho\vert$ where $\vert \nabla \rho\vert$ is the norm of $\nabla \rho$. So, the expression of the total specific energy is   
    \begin{equation}
    \alpha (\rho,\nabla\rho)=\alpha^h (\rho)+\mu\,\vert \nabla \rho\vert,
    	\label{sharp_interface_energy}
    \end{equation}
    where $\mu=const$ is a positive scalar.  The energy \eqref{sharp_interface_energy} is singular: instead of   power $2$  in  relation \eqref{energydiffuse} we have   the norm of the  density gradient in power $1$.
    \\
    We  now derive  the  equilibrium conditions for the model of volume energy.   
    We deduce the corresponding energies of a fixed volume $\mathcal D$ in the Eulerian coordinates  
    \begin{equation*}
    	E=E_1+E_2\quad{\rm with}\quad E_1=\int_{\mathcal D} \rho\, \alpha^h (\rho) \,dv \quad{\rm and}\quad E_2=\int_{\mathcal D}\mu\, \rho\,\vert \nabla \rho\vert\, dv.
    \end{equation*}
    \subsection{Variation  of $E_1$} From $\delta \rho=-\rho\, {\rm div}\,\boldsymbol{\xi}$,   we get 
    \begin{equation*}
    	\delta E_1=\int_{\mathcal D} \rho\, \delta\alpha^h (\rho) \, dv = \int_{\mathcal D} -\rho^2\, \frac{\partial\alpha^h }{\partial\rho} \, {\rm div}\,\boldsymbol \xi \, dv =-\int_{\mathcal D}{\rm div}(\mathcal P\,\boldsymbol \xi)\, dv+\int_{\mathcal D}\frac{\partial\mathcal P }{\partial\boldsymbol{x}}\,\boldsymbol \xi \, dv.
    \end{equation*}
    Due to the fact that $\boldsymbol \xi$ is zero on the boundary $\partial \mathcal D$, one has \begin{equation*}
    	\int_{\mathcal D}{\rm div}(\mathcal P\,\boldsymbol \xi)\, dv=0,
    \end{equation*} and   
    \begin{equation*}
    	\delta E_1=\int_{\mathcal D}\frac{\partial\mathcal P }{\partial\boldsymbol{x}}\,\boldsymbol \xi \, dv .
    \end{equation*}
    \subsection{Variation of $E_2$}
    \begin{equation*}
    	\delta E_2= \mu \int_{\mathcal D}\rho \,\delta\vert \nabla \rho\vert \, dv.
    \end{equation*}
    But 
    \begin{equation*}
    	\delta\vert \nabla \rho\vert =\delta\left(\frac{\partial\rho}{\partial\boldsymbol x}\right)\,\frac{\nabla \rho}{\vert \nabla \rho\vert}=\delta\left(\frac{\partial\rho}{\partial\boldsymbol x}\right)\,{\boldsymbol n},
    \end{equation*}
    where $\displaystyle {\boldsymbol n}=\frac{\nabla \rho}{\vert \nabla \rho\vert}$ is the oriented unit normal vector in increasing density to the sharp interface. From,
    \begin{equation*}
    	\delta\left(\frac{\partial\rho}{\partial\boldsymbol x}\right) = \frac{\partial\left(\delta\rho\right)}{\partial\boldsymbol x} -\frac{\partial\rho}{\partial\boldsymbol x}\,\frac{\partial\boldsymbol \xi}{\partial\boldsymbol x},
    \end{equation*}
    we get \cite{Gavrilyuk_Gouin_1999}
    \begin{equation*}
    	\delta\vert \nabla \rho\vert = -\frac{\partial (\rho\,{\rm div}\;\boldsymbol \xi)}{\partial \boldsymbol x}\,\boldsymbol n-\frac{\partial\rho}{\partial\boldsymbol x}\,\frac{\partial\boldsymbol \xi}{\partial\boldsymbol x}\, \boldsymbol n.
    \end{equation*}
    Then,
    \begin{equation*}
    	\rho\, \delta\vert \nabla \rho\vert = -\frac{\partial (\rho\,{\rm div}\;\boldsymbol \xi)}{\partial \boldsymbol x}\,\rho\,\boldsymbol n-\rho\,\frac{\partial\rho}{\partial\boldsymbol x}\,\frac{\partial\boldsymbol \xi}{\partial\boldsymbol x}\, \boldsymbol n
    \end{equation*}
    		\begin{equation*}	
    	= - \, {\rm div}\big((\rho\, {\rm div}\,\boldsymbol \xi)\,\rho\,\boldsymbol{ n}\big)+(\rho\, {\rm div}\,\boldsymbol \xi)\,{\rm div}(\rho\,\boldsymbol{ n}) -{\rm Tr\left(\rho\,\boldsymbol{n}\frac{\partial\rho}{\partial\boldsymbol x}\,\frac{\partial\boldsymbol \xi}{\partial\boldsymbol x}\right)}.
    \end{equation*}
    Let us note that 
    \begin{equation*}\displaystyle\int_{\mathcal D}{\rm div}\big((\rho\, {\rm div}\,\boldsymbol \xi)\,\rho\,\boldsymbol{ n}\big) \, dv =0,
    \end{equation*}
    \begin{equation*}
    	(\rho\, {\rm div}\,\boldsymbol \xi)\,{\rm div}(\rho\,\boldsymbol{ n})= {\rm div}\big(\rho\,{\rm div}(\rho\,\boldsymbol{n})\,\boldsymbol \xi\big) -\frac{\partial(\rho\,{\rm div}(\rho\;\boldsymbol n))}{\partial \boldsymbol x}\,\boldsymbol \xi,
    \end{equation*}
    \begin{equation*}
    	{\rm Tr\left(\rho\,\boldsymbol{n}\frac{\partial\rho}{\partial\boldsymbol x}\,\frac{\partial\boldsymbol \xi}{\partial\boldsymbol x}\right)}={\rm div}\left(\rho\,\nabla\rho\,\boldsymbol n^T\boldsymbol\xi\right)-{\rm div}\left(\rho\,\nabla\rho\,\boldsymbol n^T\right)\,\boldsymbol\xi.
    \end{equation*}
    As usual, the following integrals are vanishing at $\partial D$ 
    \begin{equation*}
    	\int_{\mathcal D}{\rm div}\big(\rho\,{\rm div}(\rho\,\boldsymbol{n})\,\boldsymbol \xi\big)\, dv =0, \quad  \int_{\mathcal D}{\rm div}\left(\rho\,\nabla\rho\,\boldsymbol n^T\boldsymbol\xi\right) \, dv=0.
    \end{equation*}
    We also have 
    \begin{equation*}
    	{\rm div}\left(\rho\,\nabla\rho\,\boldsymbol n^T\right) ={\rm div}(\rho\,\nabla\rho)\,\boldsymbol n^T+\left(\frac{\partial\boldsymbol n}{\partial\boldsymbol x}\,\boldsymbol n\right)^T \rho\,\vert \nabla \rho\vert.
    \end{equation*}
    Since  $\displaystyle {\boldsymbol n}=\frac{\nabla \rho}{\vert \nabla \rho\vert}$ , then  $\displaystyle \left(\frac{\partial\boldsymbol n}{\partial\boldsymbol x}\right)^T =\frac{\partial\boldsymbol n}{\partial\boldsymbol x}$.    It implies $\displaystyle {\rm div}\left(\rho\,\nabla\rho\,\boldsymbol n^T\right) ={\rm div}(\rho\,\nabla\rho)\,\boldsymbol n^T$. Consequently, up to conservative terms vanishing at the boundary, we abuse the notations and  write simply 
    \begin{equation*}
    	\rho\, \delta\vert \nabla \rho\vert =\left(-\frac{\partial(\rho\,{\rm div}(\rho\;\boldsymbol n))}{\partial \boldsymbol x}+{\rm div}\left(\rho\,\nabla\rho\right)\,\boldsymbol n^T\right)\;\boldsymbol \xi.
    \end{equation*}
    Using  the identities :
    \begin{equation*}
    	\frac{\partial(\rho\,{\rm div}(\rho\;\boldsymbol n))}{\partial \boldsymbol x}={\rm div}(\rho\;\boldsymbol n)  \,\frac{\partial\rho}{\partial\boldsymbol x}+\rho\,\frac{\partial({\rm div}(\rho\;\boldsymbol n))}{\partial\boldsymbol x} 
    \end{equation*}
    \begin{equation*}
    	=\frac{\partial\rho}{\partial\boldsymbol x}\,\boldsymbol n\,\frac{\partial\rho}{\partial\boldsymbol x}+\rho\,{\rm div}(\boldsymbol n)\,\frac{\partial\rho}{\partial\boldsymbol x}+ \rho\,\frac{\partial}{\partial\boldsymbol x}\left(\rho\,{\rm div}(\boldsymbol n)+\frac{\partial\rho}{\partial\boldsymbol x}\,\boldsymbol n \right)
    	  \end{equation*}
    \begin{equation*}
    	=\vert \nabla \rho\vert\,\boldsymbol n^T\boldsymbol n\, \frac{\partial\rho}{\partial\boldsymbol x}+2\,\rho\,{\rm div}(\boldsymbol n)\,\frac{\partial\rho}{\partial\boldsymbol x}+\rho^2\,\frac{\partial{\rm div}(\boldsymbol n)}{\partial\boldsymbol x}+\rho\,\frac{\partial}{\partial\boldsymbol x}\left(\frac{\partial\rho}{\partial\boldsymbol x}\,\boldsymbol n\right)
    \end{equation*}
    \begin{equation*}
    	=\vert \nabla \rho\vert\, \frac{\partial\rho}{\partial\boldsymbol x}+2\,\rho\,\vert \nabla \rho\vert\,  {\rm div}(\boldsymbol n)\,{\boldsymbol n}^T+\rho^2\,\frac{\partial{\rm div}(\boldsymbol n)}{\partial\boldsymbol x}+\rho\,\frac{\partial\vert \nabla \rho\vert}{\partial\boldsymbol x} 
    \end{equation*}
    and,
    \begin{equation*}
    	{\rm div}\left(\rho\,\nabla\rho\right)\,\boldsymbol n^T=\rho\,{\rm div}\left(\nabla\rho\right)\,\boldsymbol n^T+\frac{\partial\rho}{\partial\boldsymbol x}\,\nabla\rho\,\boldsymbol n^T
    	 \end{equation*}
     \begin{equation*}
    	= \rho\frac{\partial \vert \nabla \rho\vert}{\partial\boldsymbol x}\, \boldsymbol n \boldsymbol n^T+\rho\,\vert \nabla \rho\vert\,{\rm div}(\boldsymbol n)\,\boldsymbol n^T+\vert \nabla \rho\vert\, \frac{\partial\rho}{\partial\boldsymbol x},
    \end{equation*}
    we finally get 
    \begin{equation*}
    	\rho\, \delta\vert \nabla \rho\vert = \left( -\rho\,{\rm div}(\boldsymbol n)\,\frac{\partial\rho}{\partial\boldsymbol x}-\rho^2\frac{\partial{\rm div}(\boldsymbol n)}{\partial\boldsymbol x}-\rho\,\frac{\partial\vert \nabla \rho\vert}{\partial\boldsymbol x}+ \rho\frac{\partial \vert \nabla \rho\vert}{\partial\boldsymbol x}\, \boldsymbol n \boldsymbol n^T \right)\,\boldsymbol \xi. 
    \end{equation*}
    We deduce
    \begin{equation*}
    	\delta E = \int_{\mathcal D}\left\{ \frac{\partial\mathcal P }{\partial\boldsymbol{x}}+\mu\left(-\rho\,{\rm div}(\boldsymbol n)\,\frac{\partial\rho}{\partial\boldsymbol x}-\rho^2\frac{\partial{\rm div}(\boldsymbol n)}{\partial\boldsymbol x}+ \rho\frac{\partial \vert \nabla \rho\vert}{\partial\boldsymbol x}\, \left(\boldsymbol n \boldsymbol n^T-\boldsymbol{I}\right) \right)                    \right\}\,\boldsymbol \xi\, dv.
    \end{equation*}
    Matrix $\boldsymbol{I}$ denotes the identity tensor. Consequently, the equilibrium equation writes
    \begin{equation}
    	\frac{\partial\mathcal P }{\partial\boldsymbol{x}}-\mu\rho\left(\frac{\partial\left(\rho {\rm div}(\boldsymbol n)\right)}{\partial\boldsymbol x}+\frac{\partial \vert \nabla \rho\vert}{\partial\boldsymbol x}\, \left({\boldsymbol{I}}-\boldsymbol n \boldsymbol n^T\right) \right)              = 0.
    	\label{equilibrium_equation_sharp_interface}
    \end{equation}
    The last term in \eqref{equilibrium_equation_sharp_interface} contains only tangential derivatives  of the density along each surface  $\rho=const$ (which is equivalent to  $\mathcal P =const$) defining the interface.  From \eqref{equilibrium_equation_sharp_interface}, we obtain,  
    \begin{equation}
    	\frac{\partial\mathcal P }{\partial\boldsymbol{x}}-\mu\,\rho\left(-H_s\frac{\partial\rho }{\partial\boldsymbol x}+\frac{\partial \vert \nabla \rho\vert}{\partial\boldsymbol x}\, \left({\boldsymbol{I}}-\boldsymbol n \boldsymbol n^T\right) \right)              = 0.
    	\label{equilibrium_equation_sharp_interface_2}
    \end{equation}
    The equilibrium condition \eqref{equilibrium_equation_sharp_interface_2} should be satisfied in the domains outside the interface separating different phases. \\  
    In curvilinear coordinates defined in Section 1, we get the following relations on the  surface 
    \begin{equation*}
    	\vert \nabla \rho\vert=\boldsymbol e_3^T\,\nabla\rho=\frac{1}{h_3}\frac{\partial \rho}{\partial q_3}
    	\end{equation*}
    	and
    	   \begin{equation*}
 \frac{\partial \vert \nabla \rho\vert}{\partial\boldsymbol x}\, \left({\boldsymbol{I}}-\boldsymbol n \boldsymbol n^T\right)
=
\frac{1}{h_3}\frac{\partial}{\partial q_3}\left(\frac{1}{h_3}\frac{\partial\rho}{\partial q_3}\right){\boldsymbol e}_3^T\left({\boldsymbol{I}}-\boldsymbol e_3 \,\boldsymbol e_3^T\right)=0.
 \end{equation*}
    Then, \eqref{equilibrium_equation_sharp_interface_2} yields
    \begin{equation}
    	\nabla \mathcal P+\mu \,\rho\,H_s\,\frac{1}{h_3}\frac{\partial \rho}{\partial q_3}\, \boldsymbol e_3=0.
    	\label{equilibrium_equation_sharp_inter}
    \end{equation}
    Along the interface, the  pressure is constant 
    \begin{equation*}
    	\nabla_{tg} \mathcal P=0.
    \end{equation*}
    In projection along $\boldsymbol n=\boldsymbol e_3$ the  equilibrium equation \eqref{equilibrium_equation_sharp_inter} writes
    \begin{equation}
    	\frac{\partial\mathcal P }{\partial{x_3}}+\mu\, \rho\,H_s\,\frac{\partial\rho}{\partial x_3}              = 0.\label{equilibrium_equation_sharp}
    \end{equation}
    By integrating \eqref{equilibrium_equation_sharp} from liquid to vapor bulks, we obtain
    \begin{equation*}
    	\mathcal P_v-\mathcal P_l=\frac{\mu}{2}\,(\rho_l^2-\rho_v^2)\,H_s.
    \end{equation*} 
    Hence, we obtain the Laplace law 
    \begin{equation}
    	\mathcal{P}_\ell-\mathcal{P}_v= \sigma \, H_s,
    	\label{Laplace_law_2}
    \end{equation}
    where the surface tension coefficient $\sigma$ is 
    \begin{equation}
    	\displaystyle\sigma =\frac{\mu}{2}\,(\rho_l^2-\rho_v^2).
    	\label{definition2_sigma}	
    \end{equation}
    For a given surface tension value, relation  \eqref{definition2_sigma}	can be considered as the definition of parameter  $\mu$. 
    \section{Comparison of Laplace's law between diffuse and sharp interfaces}
    The Laplace's laws \eqref{Laplace_law_1} \eqref{Laplace_law_2}  are written in a similar way.  However, in the case of diffuse interfaces, the surface tension coefficient is defined by \eqref{definition1_sigma} and determined by the whole density profile (and corresponds to the  case of \textit{thick interfaces}). In the case of \textit{sharp interfaces}, there is no   density profile through  the geometric interface, the  surface tension coefficient is given in terms of   jump of density. The  formula \eqref{definition2_sigma}	can be interpreted  as follows. Let 
    	\begin{equation*}
    		E_2=\mu\int_{\mathcal D} \rho\,\vert \nabla \rho\vert\, dv.
    	\end{equation*}
    	when $\mathcal D = \Sigma_\rho\times [\rho_v,\rho_\ell]$, where $\Sigma_\rho$ is the isodensity surface associated with $\rho$, with $\rho \in [\rho_v,\rho_\ell]$.
    	Let us consider a regular function $\phi(\boldsymbol x)$ with compact support in $\mathcal D$. One defines the  distribution $\rho\,\vert \nabla \rho\vert$ (see \cite{Schwartz_1966})
    	\begin{equation*}
    		<\rho\,\vert \nabla \rho\vert,\phi> =
    		 \frac{1}{2}\int_{\mathcal D}\,\phi\,{\rm det}\left( \boldsymbol n,d_1\boldsymbol x,d_2\boldsymbol x\right)\,\frac{d (\rho^2)}{d\ell}\, d\ell= \frac{1}{2}\int_{\rho_v}^{\rho_\ell}\,\left(\int_{\Sigma_\rho}\phi\,ds\right)\,\frac{d (\rho^2)}{d\ell}\, d\ell,
    	\end{equation*}
    	Here $d_1\boldsymbol x$ and $d_2\boldsymbol x$ are two  convenient infinitesimal  tangent vectors to $\Sigma_\rho$, and $ds$ is an infinitesimal surface element. 
    	At  the limit when the surfaces $\Sigma_\rho$ approach the surface $\Sigma$ representing an interface of zero thickness,
    	\begin{equation}
    		<\rho\,\vert \nabla \rho\vert,\phi>\equiv\int_{\mathcal D} \rho  \vert \nabla \rho\vert\, \phi\, dv \, \longrightarrow\, \int_{\Sigma }  \left[\frac{\rho^2}{2}\right]\,\phi\, ds,
    		\label{limit_formula}
    	\end{equation}
     where\,\ $\displaystyle \left[\frac{\rho^2}{2}\right]=\frac{\rho_\ell^2-\rho_v^2}{2}$.\\
    Introducing the \textit{surface delta function}  usually denoted $\delta_\Sigma$ 
    	\begin{equation*}
    		<\delta_\Sigma,\phi> =\int_{\Sigma }   \phi \, ds.
    	\end{equation*}
    	and $g(x)$ for the constant jump of $\rho^2/2$, we interpret the  formula \eqref{limit_formula} as
    	\begin{equation*}
    		<g(x)\delta_\Sigma,\phi> =\int_{\Sigma }  g(x) \phi \, ds.
    	\end{equation*}
	\bibliographystyle{amsrefs}
 \bibliography{references_CR}
\end{document}